\documentclass[11pt]{article}
\usepackage[dvips]{graphicx}
\usepackage{color}

\begin {document}

\definecolor{shade}{gray}{0.6}

\begin{center}
\begin{Large}
\bf{NEW MEASUREMENTS OF THE $\eta$ and $K^0$ MASSES}
\end{Large}
\end{center}
\begin{center}
\small{
The NA48 Collaboration  \\
\begin{center}
\small{
A.~Lai, D.~Marras \\
{\small \em Dipartimento di Fisica dell'Universit\`a e Sezione
    dell'INFN di Cagliari, I-09100 Cagliari, Italy} \\[0.35cm]

J.R.~Batley,
A.J.~Bevan\footnote{
Present address: Oliver Lodge Laboratory,
University of Liverpool, Liverpool L69 7ZE, U.K.},
R.S.~Dosanjh,
T.J.~Gershon\footnote{
Present address: High Energy Accelerator Research
Organization (KEK), Tsukuba, Ibaraki, 305-0801, Japan},
G.E.~Kalmus, C.~Lazzeroni, D.J.~Munday, E.~Olaiya,
M.A.~Parker, T.O.~White, S.A.~Wotton \\
{\small \em Cavendish Laboratory, University of Cambridge,
    Cambridge, CB3 0HE, UK\/}\footnote{Funded by the UK
    Particle Physics and Astronomy Research Council} \\[0.35cm]

G.~Barr,
G.~Bocquet, A.~Ceccucci, T.~Cuhadar-D\"{o}nszelmann, D.~Cundy,
G.~D'Agostini, N.~Doble, V.~Falaleev, W.~Funk, L.~Gatignon, A.~Gonidec,
B.~Gorini, G.~Govi, ,P.~Grafstr\"om, W.~Kubischta, A.~Lacourt,
M.~Lenti\footnote{
On leave from Sezione dell'INFN di Firenze,
I-50125, Firenze, Italy},
I.~Mikulec\footnote{
On leave from \"Osterreichische Akademie der
Wissenschaften, Institut f\"ur Hochenergiephysik, A-1050 Wien, Austria},
A.~Norton, S.~Palestini, B.~Panzer-Steindel, D.~Schinzel,
G.~Tatishvili\footnote{
On leave from Joint Institute for Nuclear Research,
Dubna, 141980, Russian Federation},
H.~Taureg, M.~Velasco, H.~Wahl \\
{\small \em CERN, CH-1211 Geneva 23, Switzerland} \\[0.35cm]

C.~Cheskov
P.~Hristov,
V.~Kekelidze,
D.~Madigojine,
N.~Molokanova,
Yu.~Potrebenikov,
A.~Zinchenko \\
{\small \em Joint Institute for Nuclear Research, Dubna, Russian
    Federation}\\[0.35cm]

V.J.~Martin\footnote{
Permanent address: Department of Physics and
Astronomy, Northwestern
University, 2145 Sheridan Road, Evanston IL 60202, USA},
P.~Rubin\footnote{
Permanent address: Department of Physics,
University of Richmond, VA 27313, USA},
R.~Sacco,
A.~Walker \\
{\small \em Department of Physics and Astronomy, University of
    Edinburgh, JCMB King's Buildings, Mayfield Road, Edinburgh,
    EH9 3JZ, UK\/}$^{3)}$ \\[0.35cm]

D.~Bettoni, R.~Calabrese,
M.~Contalbrigo, P.~Dalpiaz, J.~Duclos,
P.L.~Frabetti\footnote{
Permanent address: Dipartimento di
Fisica dell'Universit\`a e Sezione dell'INFN di Bologna,
I-40126 Bologna, Italy},
A.~Gianoli, E.~Luppi,
M.~Martini,
L.~Masetti, F.~Petrucci, M.~Savri\'e, 
M.~Scarpa \\
{\small \em Dipartimento di Fisica dell'Universit\`a e Sezione
    dell'INFN di Ferrara, I-44100 Ferrara, Italy} \\[0.35cm]

A.~Bizzeti\footnote{
Also at Dipartimento di Fisica
dell'Universit\`a di Modena, I-41100 Modena, Italy},
M.~Calvetti, G.~Collazuol,
G.~Graziani, E.~Iacopini,
F.~Martelli\footnote{
Instituto di Fisica Universit\'a di Urbino},
M.~Veltri$^{11)}$ \\
{\small \em Dipartimento di Fisica dell'Universit\`a e Sezione
    dell'INFN di Firenze, I-50125 Firenze, Italy} \\[0.35cm]

H.G.~Becker,
M.~Eppard, H.~Fox, A.~Hirstius, K.~Holtz,
A.~Kalter, K.~Kleinknecht, U.~Koch, 
L.~K\"opke,
P.~Lopes~da~Silva, P.~Marouelli, I.~Mestvirishvili,
I.~Pellmann, A.~Peters,
S.A.~Schmidt,
V.~Sch\"onharting, Y.~Schu\'e,
R.~Wanke, A.~Winhart, M.~Wittgen \\
{\small \em Institut f\"ur Physik, Universit\"at Mainz, D-55099
    Mainz, Germany\/}\footnote{Funded by the German Federal Minister for
    Research and Technology (BMBF) under contract 7MZ18P(4)-TP2} \\[0.35cm]

J.C.~Chollet, L.~Fayard, 
L.~Iconomidou-Fayard,
J.~Ocariz,
G.~Unal, I.~Wingerter-Seez \\
{\small \em Laboratoire de l'Acc\'{e}l\'{e}rateur Lin\'{e}aire, IN2P3-CNRS,
Universit\'{e} de Paris-Sud, \\
F-91405 Orsay, France\/}\footnote{Funded by Institut National de Physique
des Particules et de Physique Nucl\'{e}aire (IN2P3), France} \\[0.35cm]

G.~Anzivino, P.~Cenci, E.~Imbergamo,
G.~Lamanna,
P.~Lubrano, A.~Mestvirishvili, A.~Nappi,
M.~Pepe, M.~Piccini, \\
{\small \em Dipartimento di Fisica dell'Universit\`a e Sezione
    dell'INFN di Perugia, I-06100 Perugia, Italy} \\[0.35cm]

R.~Casali, C.~Cerri,
M.~Cirilli\footnote{
Present address: Dipartimento di Fisica dell'Universit\'a
di Roma ``La Sapienza'' e Sezione INFI di Roma, I-00185 Roma, Italy},
F.~Costantini, R.~Fantechi, 
L.~Fiorini, S.~Giudici,
I.~Mannelli,
G.~Pierazzini,
M.~Sozzi \\
{\small \em Dipartimento di Fisica dell'Universit\`a,
Scuola Normale Superiore\\
e Sezione dell'INFN di Pisa, I-56100 Pisa, Italy} \\[0.35cm]

J.B.~Cheze, J.~Cogan, M.~De Beer, 
P.~Debu, F.~Derue, A.~Formica,
G.~Gouge, R.~Granier de Cassagnac, G.~Marel,
E.~Mazzucato, B.~Peyaud, R.~Turlay, 
B.~Vallage \\
{\small \em DSM/DAPNIA - CEA Saclay, F-91191 Gif-sur-Yvette,
    France} \\[0.35cm]

M.~Holder, A.~Maier, M.~Ziolkowski \\
{\small \em Fachbereich Physik, Universit\"at Siegen, D-57068
Siegen, Germany\/}\footnote{Funded by the German Federal Minister for
Research and Technology (BMBF) under contract 056SI74} \\[0.35cm]

R.~Arcidiacono, C.~Biino, 
N.~Cartiglia, R.~Guida,
F.~Marchetto, E.~Menichetti,
N.~Pastrone \\
{\small \em Dipartimento di Fisica Sperimentale dell'Universit\`a e
    Sezione dell'INFN di Torino, \\ I-10125 Torino, Italy} \\[0.35cm]

J.~Nassalski, E.~Rondio, 
M.~Szleper, W.~Wislicki, S.~Wronka \\
{\small \em Soltan Institute for Nuclear Studies, Laboratory for High
    Energy Physics, \\ PL-00-681 Warsaw, Poland\/}\footnote{
   Supported by the Committee for Scientific Research grants
   5P03B10120, 2P03B11719 and SPUB-M\/CERN\/P03\/DZ210\/2000
    and using computing resources of the Interdisciplinary Center for
    Mathematical and
    Computational Modelling of the University of Warsaw} \\[0.35cm]

H.~Dibon, M.~Jeitler, M.~Markytan, 
G.~Neuhofer,
M.~Pernicka, A.~Taurok, L.~Widhalm \\
{\small \em \"Osterreichische Akademie der Wissenschaften, Institut
    f\"ur Hochenergiephysik, \\ A-1050 Wien, Austria\/}\footnote{
    Funded by the Austrian Ministry for Traffic and Research under the
   contract
   GZ 616.360/2-IV GZ 616.363/2-VIII, Austria and by the Fonds f\"ur
    Wissenschaft und Forschung FWF Nr. P08929-PHY} \\[0.35cm]

}
\end{center}

}
\end{center}


\begin{abstract}\noindent
New measurements of the $\eta$ and $K^0$ masses have been performed
using decays to 3$\pi^0$ with the NA48 detector at the CERN SPS.
Using symmetric decays to reduce systematic effects, the results
$M(\eta) = 547.843\pm0.051$~MeV/c$^2$ and
$M(K^0) = 497.625\pm0.031$~MeV/c$^2$ were obtained.
\end{abstract}


\section*{Introduction}

\setcounter{footnote}{0}

 Precise values of the $\eta$ and $K^0$ masses are often used as input
for measurements. For instance, $\eta$ decays involving photons are 
used 
for the precise in situ calibration of electromagnetic calorimeters.
One such example is the measurement of the CP violating quantity
$\epsilon'/\epsilon$ performed by NA48~\cite{na48_epsilon}. The
current world average~\cite{pdg} relative uncertainty on the $K^0$ mass is
$\pm6\times 10^{-5}$, dominated by a measurement at a $\phi$ factory
using $\pi^+\pi^-$ decays~\cite{kl_novosib}. The $\eta$ mass is
less well known, the current uncertainty being $\pm2\times 10^{-4}$. The
most precise results in this case come from
measurements of the production cross-section
near threshold in the reactions
$d p \rightarrow \eta ^3He$ ~\cite{eta_cross} and
in $\gamma p \rightarrow \eta p$ ~\cite{eta_cross2}.
 
 In this paper, we present new measurements of the $\eta$ and $K^0$
masses using decays to 3$\pi^0$ ($\rightarrow 6 \gamma$), with
the NA48 experiment at the CERN SPS. The paper is organised as
follows: the first section describes the method used for the 
reconstruction of the $\eta$ and
$K^0$ masses; the second section shows the experimental
setup; details of the performances of the electromagnetic calorimeter
are given in the third section; results and cross-checks with a detailed
evaluation of the systematic effects are presented in the fourth section.

\section{Method}

 This measurement uses $\eta$ and $K_L$ particles
with average energies of $\approx$ 110 GeV.
Photons from 3$\pi^0$ decays are detected with a precise quasi-homogeneous
liquid krypton calorimeter. This device allows 
photon energies as well as impact positions in the plane
orthogonal to the $\eta$ and $K_L$ momentum
to be measured accurately. Although photon angles are not
measured directly with the calorimeter, the decay 
position can be inferred from two photons coming from a $\pi^0$ decay
using the $\pi^0$ mass constraint. In the limit of small opening
angles (valid to better than $10^{-5}$ for this experiment), the
distance $d$ between the decay position and the calorimeter 
nominal position can
be computed as
$$ d = \frac{1}{M_{\pi^0}} \sqrt{E_1 E_2} d_{12} $$
where $M_{\pi^0}$ is the $\pi^0$ mass, known with  $4\times10^{-6}$
accuracy\cite{pdg},
$E_1$ and $E_2$ are
the energies of the two photons from a $\pi^0$ decay and $d_{12}$ is
the distance between the two photons in the plane orthogonal to the
beam axis at the calorimeter nominal position. 
Using the average value $d_{\pi^0}$
of $d$ from the
three $\pi^0$ mass constraints, the full 6-body invariant mass can then
be computed as
$$ M = \frac{1}{d_{\pi^0}} \sqrt{\Sigma_{ij,i<j} E_i E_j d_{ij}^2} $$

 The advantages of using this method for the mass measurement are the
following:
\begin{itemize}
\item[-]{The 3$\pi^0$ decay mode is virtually background free.}
\item[-]{Thanks to the $\pi^0$ mass constraint, the resolution on $M$ is
better than 1 MeV/c$^2$.}
\item[-]{The measurement of $M$ is independent of the
energy scale of the calorimeter. Indeed, the measured quantity is
the ratio of the $\eta$ (or $K^0$) mass to the $\pi^0$ mass, in which
the absolute energy scale completely cancels. For the same reason,
the absolute transverse size scale also cancels.}
\item[-]{The measurement is only sensitive to residual non-linearities
in the energy or position measurements. To minimise the sensitivity
to energy non-linearities, only symmetric decays are used, in which the
photons all have about the same energy. Indeed for perfectly
symmetric decay, the sensitivity to energy non-linearities also cancels
.}
\item[-]{The same method can be applied to both the $\eta$ and the $K_L$
cases. The very large statistics available in the $K_L$ sample allows
many systematic cross-checks to be performed to validate the measurement.}
\end{itemize}

 The mass measurement relies only on measurements of relative
distance between photons and ratio of energies with the calorimeter.
Furthermore the performance of this device can be studied in detail 
in situ using $K_{e3}$ ($K_L \rightarrow
\pi^{\pm} e^{\mp} \nu$) decays, 
and $\pi^0$ and $\eta$ decays to two photons,
which are present with very large statistics in the data samples.

\section{Beam and detectors}

 The NA48 experiment is designed for a precise measurement of
the quantity $\epsilon'/\epsilon$ in neutral kaon decays~\cite{na48_epsilon}.
In the following only the parts of the apparatus relevant for the
measurements of the $\eta$ and $K^0$ masses are described.

\subsection{The $K_L$ beam}

 The neutral $K_L$ beam~\cite{beam} is derived from 450~GeV/c protons extracted
from the CERN SPS. For each SPS pulse (2.4~s spill every 14.4~s),
$1.5\times10^{12}$ protons hit a 40~cm long beryllium target, with
an incidence angle of 2.4~mrad. The charged component of the outgoing
particles is swept away by bending magnets. A neutral secondary
beam with a $\pm$0.15~mrad divergence is generated using three stages
of collimation over 126~m, the final element being located just upstream
of the beginning of the decay region. 
Because of the long distance between the target and the beginning
of the decay region, the decays in this neutral beam are
dominated by $K_L$. The decay region is contained in
an evacuated ($<$ $3\times10^{-5}$~mbar) 90~m long tank. During the
1999 data taking, this decay region was terminated by a polyimide
(Kevlar) composite window 0.9~mm (0.003~radiation lengths) thick.
 Downstream of it a 16~cm diameter evacuated tube passes
through all detector elements to
allow the undecayed neutral beam to travel in vacuum
all the way to a beam dump.
 In the normal $\epsilon'/\epsilon$ data taking, a $K_S$ beam is
also created using a second target closer to the decay region. The
beam from which the decays originate is identified by tagging
the protons directed to the $K_S$ target.

\subsection{The $\eta$ beam}

 During special data taking periods, a beam of negatively charged
secondary particles is, by a suitable choice of the magnets
in the beam line, directed 
along the nominal $K_L$ beam axis. This beam consists
mostly of $\pi^-$ with a broad momentum spectrum of
average energy$\approx100$~GeV,
and with a flux of $\approx1.3\times10^6$ per
pulse. It impinges on two thin polyethylene targets
(2.0~cm thick), separated by 1462~cm, near the beginning
and towards the end of the fiducial Kaon decay region. Through charge exchange
reactions,  $\pi^0$ and $\eta$ are created and,
because of their very short lifetimes,
decay inside the targets.
Neutral Kaon are also produced in the targets,
mostly from $K^-$ interactions.

\subsection{The detector}

\begin {figure}[ht]
\begin{center}
\includegraphics*[scale=0.45]{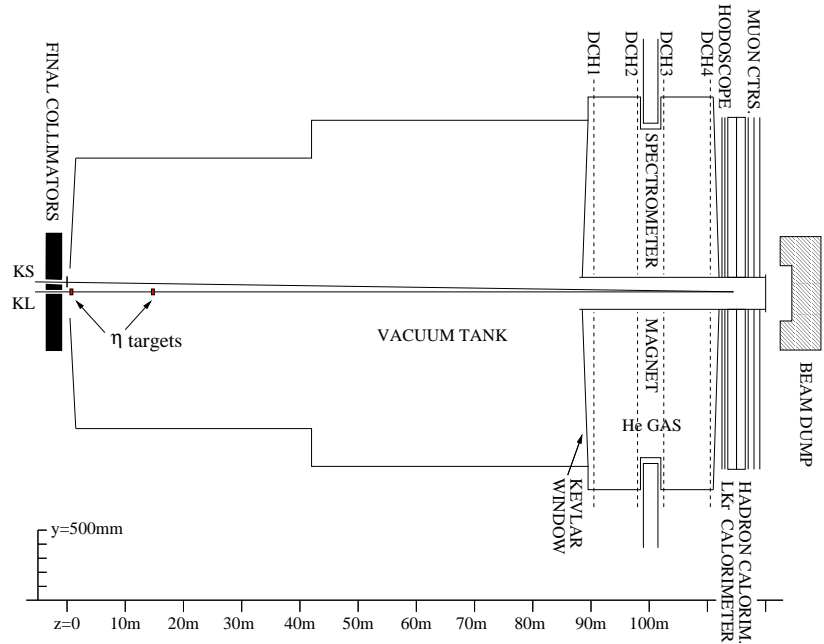}
\caption {Layout of the main detector components.}
\label {fig:detector.eps}
\end{center}
\end {figure}

 The layout of the detector is shown in Figure~\ref{fig:detector.eps}.
A magnetic spectrometer~\cite{spectro},
housed in a helium tank,
is used to measure
charged particles. It is comprised of four drift chambers and
a dipole magnet, located between the second and the third drift chambers,
giving a horizontal transverse momentum kick of 265 MeV/c. 
The momentum resolution is $\sigma_p/p = 0.5\% \oplus 0.009\% \times p$
($p$ in GeV/c), where $\oplus$ means that the contributions should be
added in quadrature.
Two plastic scintillator hodoscope planes are placed after the helium tank.

 A quasi-homogeneous liquid krypton electromagnetic calorimeter (LKr) is used
to measure photon and electron showers. It consists of $\approx$10~m$^3$
of liquid krypton with a total thickness of 127~cm
($\approx$ 27 ~radiation lengths) and an octagonal shaped
cross-section of $\approx$5.5~m$^2$. The entrance face of the
calorimeter is located 11501~cm from the beginning of the
decay region and 11455~cm from the location of the first $\eta$ target.
 Immersed in the krypton are electrodes made
with Cu-Be-Co ribbons
of cross-section $40 \mu$m$\times 18 $mm$\times 127 $cm, extending
between the front and the back of the detector.
The ribbons are guided through precisely
machined slots in five spacer plates located longitudinally every 21~cm
which ensure the gap width stability in the shape of a $\pm$~48~mrad
accordion.
The 13212 readout cells each have a cross-section of 2$\times$2~cm$^2$
at the back of the active region,
and consist (along the horizontal direction)
of a central anode at a 3 kV high voltage in the middle
of two cathodes kept at ground.
The cell structure is projective towards
the middle of the decay region, 110~m upstream of the calorimeter, so that
the measurement of photon positions is insensitive to first order to the
initial conversion depth. 
The initial current induced on the electrodes by the drift of the ionization
is measured using pulse shapers with 80~ns FWMH and digitised with 40~MHz
10-bits FADCs~\cite{lkr_elec}. The dynamic range is increased by employing
four gain-switching amplifiers which change the amplification factor
depending on the pulse height. The electronic noise per cell
is $\approx$ 10 MeV. 
The calorimeter is housed in a vacuum insulated cryostat. The total
amount of matter before the beginning of the liquid krypton volume
is $\approx$ 0.8 radiation lengths, including all the NA48 detector
elements upstream of the calorimeter.
The total output of the calorimetric information is restricted
to cells belonging to a region around the most energetic cells,
by a ``zero-suppression'' algorithm. For each event, 
the information from 10 digitised time samples (250~ns) is read out.
 An iron-scintillator hadron calorimeter is located downstream of the LKr,
followed by a muon veto system. 

\subsection{Triggers and data sets}
 
 The trigger on multiphoton final states~\cite{nut} 
operates on the analogue sums
of signals from $2\times8$ cells of the LKr calorimeter in both
horizontal ($x$) and vertical ($y$) orientations. The signals are digitised
and summed in $x$ and $y$ projections. Based on these projections, the
total deposited energy, the energy-weighted centre of gravity
of the event and the estimated decay vertex position along
the beam axis are computed.

 To select $K_L \rightarrow 3\pi^0$ decays, the total energy is required
to be more than 50 GeV, the radial position of the centre of gravity to
be less than 15~cm from the beam axis
and the decay vertex position to be less than
9 $K_S$-lifetimes away from the beginning of the decay region.

 To select $\pi^0$ and $\eta$ decays in the $\eta$ runs, a cut
on the total energy is applied. Events with energy above 90 GeV are
taken without downscaling, events with energy between 40 and 90 GeV
are downscaled by a factor 10 or 12 depending on the run period, and
events with energy between 15 and 40 GeV are downscaled by a factor 4 to 6.

 During the $\epsilon'/\epsilon$ run in 1998 and 1999, large
statistics of $K_{e3}$ decays were collected to study the calorimeter
performance.
 In the year 2000, the detector operated with vacuum in place
of the spectrometer. During this period, data with only the $K_L$ beam
were recorded, as well as data with an $\eta$ beam. These data are used
for the mass measurements presented here.

\section{Performance of the electromagnetic calorimeter}

\subsection{Reconstruction}

 Photon or electron showers are found in the calorimeter by looking
for maxima in the digitised pulses from individual cells in both
space and time. Energy and time of the pulses are estimated using
a digital filter technique. The first calibration is performed using
a calibration pulser system. Small drifts of the pedestal due to
temperature effects are monitored and corrected for. The energy
of the shower is computed accumulating the energy in the cells
within a radius of 11~cm, containing more than 90\%
of the shower total energy (this fraction being
constant with energy). The shower position is derived from the
energy-weighted centre of gravity of the 3$\times$3 central cells,
corrected for the residual bias of this estimator.
Shower energies are corrected for the following effects:
small ($<$ 1\%) variations of the energy measurement depending on the impact
point within the cell, energy lost outside the calorimeter
boundaries, energy lost in non-working cells (about 0.4\% of the
channels), average energy lost in the material before the calorimeter
(15~MeV for photons, 45~MeV for electrons), a small bias from
the zero-suppression algorithm for energies below 5 GeV and 
small space charge effects from the accumulation of positive ions
in the gap~\cite{space_charge}
which is significant only
in the case of data taken with the high intensity $K_L$
beam.
 Energy leakage from one shower to another is corrected in
the reconstruction using the transverse shower profile from a
GEANT~\cite{geant} simulation. Special runs in which a monochromatic
electron beam is sent into the calorimeter, taken
without zero suppression, are used to check this modelling and
to derive a small correction.

Fiducial cuts are applied to ensure that shower energies are
well measured. These cuts include the removal of
events with a shower too close
to the inner beam tube or the outer edges of the LKr or falling
within 2~cm of a non-working cell. Only
showers in the energy range 3-100 GeV are used.

The corrections described above are the same as those
applied for the $\epsilon'/\epsilon$ measurement discussed
in \cite{na48_epsilon}.

\subsection{Performance from $K_{e3}$ decays}

 More than $150\times10^6$ $K_{e3}$ decays have been
recorded during the  $\epsilon'/\epsilon$ data taking periods. The comparison
of the electron momentum measured by the magnetic spectrometer ($p$)
and the energy measured by the LKr ($E$) allows 
the performance of the calorimeter to be studied in detail, 
the resolution on the ratio
$E/p$ being $\approx$ 1\%.
 The $K_{e3}$ sample is first used to study and improve the cell-to-cell
intercalibration of the calorimeter. After the electronic calibration,
the cell-to-cell dispersion in the shower energy measurement is about 0.4\%.
Electrons from $K_{e3}$ in the energy range 25-40 GeV (to avoid
possible correlations between non-uniformity and non-linearity) are
used to equalise the cell-to-cell energy response, requiring $E/p$
to be the same for all cells. This is done with a statistical accuracy of
about 0.15\% using the 1998 data sample. Residual long range variations 
of the energy response over
the calorimeter are expected to be smaller.

\begin {figure}[ht]
\begin{center}
\includegraphics*[scale=0.45]{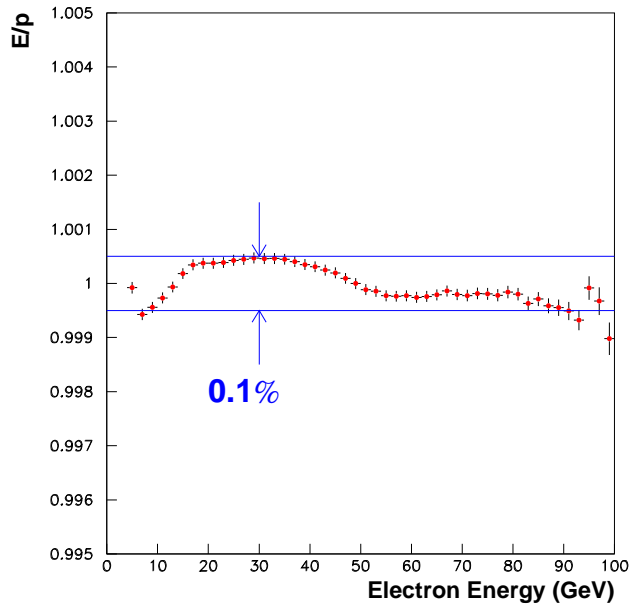}
\caption {Limit to the non-linearity of energy measurement from $K_{e3}$ decays}
\label {fig:ep.eps}
\end{center}
\end {figure}

 Unfolding the momentum resolution from the measured $E/p$ resolution
as a function of the electron energy, the LKr energy resolution
can be obtained:
$$\frac{\sigma(E)}{E} = \frac{(0.032\pm0.002)}{\sqrt{E}} \oplus \frac{(0.090\pm0.010)}{E} \oplus (0.0042\pm0.0005) $$
where $E$ is the energy in GeV.
The sampling term is dominated by fluctuations in the shower leakage
outside the area used to measure the energy. The constant term arises
from residual variations with the impact point, accuracy in the
pulse reconstruction, residual gap variations and inter-cell calibration.
 The energy response is linear within about 0.1\% over the energy
range 5-100 GeV, as illustrated in Figure~\ref{fig:ep.eps} which shows
the average value of $E/p$ as a function of the electron energy. The
measured small variation of $E/p$ as a function
of energy is applied as a correction to the photon energies.

 The comparison of the reconstructed electron position using the calorimeter
with that of the extrapolated track allows the position measurement
of the LKr to be adjusted and its transverse size scale to
be checked with an accuracy of
$2.5\times10^{-4}$. The position resolution 
of the calorimeter is better than 1.3~mm
above 20 GeV.

\subsection{Study of systematic effects using $\pi^0$ decays}

\begin {figure}[ht]
\begin{center}
\includegraphics*[scale=0.45]{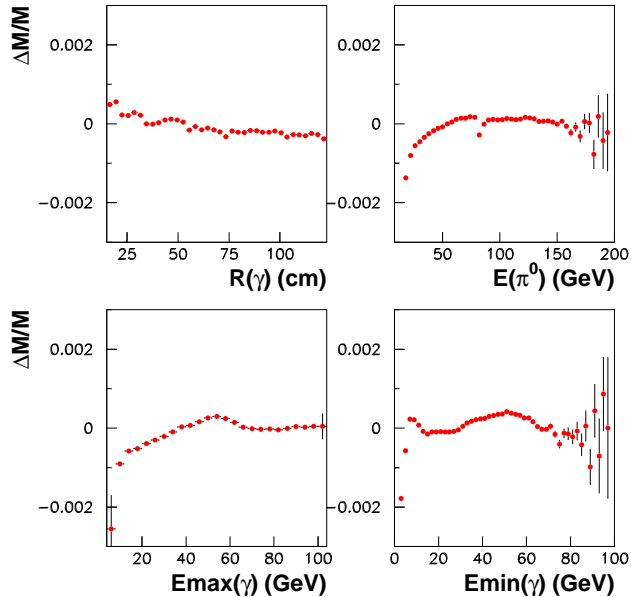}
\caption {Relative variation of the reconstructed $\pi^0$ mass, as a function
of the impact radius of photons, total $\pi^0$ energy and maximum and
minimum energy of the two photons}
\label {fig:pi0_gg.eps}
\end{center}
\end {figure}

 Using the large sample of $\pi^0$ decays to two photons
produced during the $\eta$ runs, the
systematic effects in the calorimeter measurement can be estimated.
From the known target positions, the $\pi^0$ mass can be reconstructed
from the measured photon energies and positions at the LKr.
Figure~\ref{fig:pi0_gg.eps} shows the relative variation
of the reconstructed $\pi^0$ mass as a function of several variables:
photon impact position, $\pi^0$ energy, maximum and minimum photon
energy. Significant non-uniformities in the energy response would lead
to a visible variation of the $\pi^0$ mass with the photon impact point.
Non-linearities in the energy response would lead to drifts in
the reconstructed mass as a function of the energy.
As can be seen in Figure~\ref{fig:pi0_gg.eps}, the $\pi^0$ mass is stable
to $\approx$ 0.1\%, except in the energy region $E_{\gamma} < 6$ GeV, where
stronger effects can be seen. However, this energy region is not
used in the analysis of symmetric 3$\pi^0$ decays, where the
photon energies are in the 8 to 37 GeV range.
The $\pi^0$ sample is also used for the 
2000 data to improve the 1998  $K_{e3}$ intercalibration,
by equalising the reconstructed $\pi^0$ mass
value as a function of the photon impact cell. The dispersion of the correction
coefficients is about 0.1\%.

The non-linearities in the energy response can be parameterised as:
\begin{equation}
\frac{\Delta E}{E} = \frac{\alpha}{E} + \beta E + \gamma r 
\end{equation}
where $E$ is the photon energy and $r$ the impact radius over the
calorimeter.
From the $\pi^0 \rightarrow \gamma \gamma$ data,
$\alpha$ can be constrained to be within $\pm$ 10 MeV,
$\beta$ within
$\pm 2\times10^{-5}$ GeV$^{-1}$ and $\gamma$ within $\pm 10^{-5}$ cm$^{-1}$.

\section{$\eta$ and $K^0$ masses measurement}

\subsection{Selection of 3$\pi^0$ decays}

 To select 3$\pi^0$ decays, any group of six showers within
$\pm 5$~ns of their average time is examined. No other
shower above 1.5 GeV should be present within a $\pm$ 3~ns time.
The minimum distance between photon candidates is required
to be greater than 10~cm. The total energy is required
to be in the range 70-170 GeV for $K_L$, and 70-180 GeV for $\eta$
decays. Using the six photons, the decay vertex position along
the beam axis
is computed assuming the $K^0$ or the $\eta$ mass
\footnote{The assumed mass values at this stage are only used to perform a loose
selection, and they do not affect the final mass measurement}. 
The resolution
is $\approx$ 50 cm. In the $\eta$ case, this position is required
to be consistent with the position of one of the two targets
within 400~cm. In the $K_L$ case,
the decay vertex position should be
less than 6.5 $K_S$ lifetimes downstream of the
beginning of the decay region.
Using this decay vertex, the invariant mass of
any pair of photons is computed. A $\chi^2$ like variable is
computed comparing the masses of the three pairs to the nominal
$\pi^0$ mass. The typical
$\pi^0$ mass resolution is around 1 MeV and is
parametrised as a function of the lowest energy photon.
The pairing with the smallest $\chi^2$ is selected.
To remove any residual background (from events with
a $\pi^0$ Dalitz decay for example), a loose $\chi^2$ cut is applied.
The correct $\pi^0$ pairing is selected by this method in
99.75\% of the events, and the residual bias 
on the mass result induced by wrong pairing
is completely negligible.
This procedure selects $128\times10^6$ $K_L \rightarrow 3\pi^0$ candidates
and $264\times10^3$ $\eta \rightarrow 3\pi^0$ candidates in the data from the
year 2000.

 To minimise the sensitivity of the measurement to residual
energy non-linearity, configurations in which
the photons have comparable energies are selected using
the following cut on each photon energy:
$$ 0.7 < \frac{E_{\gamma}}{\frac{1}{6} E_{tot}} < 1.3 $$
 where $E_{\gamma}$ is the energy of the photon and $E_{tot}$ the
sum of the energies of all photons.
 This cut leaves a sample of $655\times10^3$ $K_L$ decays and 
1134 $\eta$ decays.
The events are then subjected to the mass reconstruction described
in section 1. 
In these computations, photon positions are evaluated
at the position of the maximum of the shower in the calorimeter, to
account correctly for deviations of photon directions from the
projectivity of the calorimeter.
The same procedure is applied to Monte-Carlo
samples of $K_L$ and $\eta$ decays, following the same
reconstruction and analysis path, to check for any bias in the
procedure. The simulation of the calorimeter response is based on a large
shower library generated with GEANT.

\subsection{Results and cross-checks}

\begin {figure}[ht]
\begin{center}
\includegraphics*[scale=0.45]{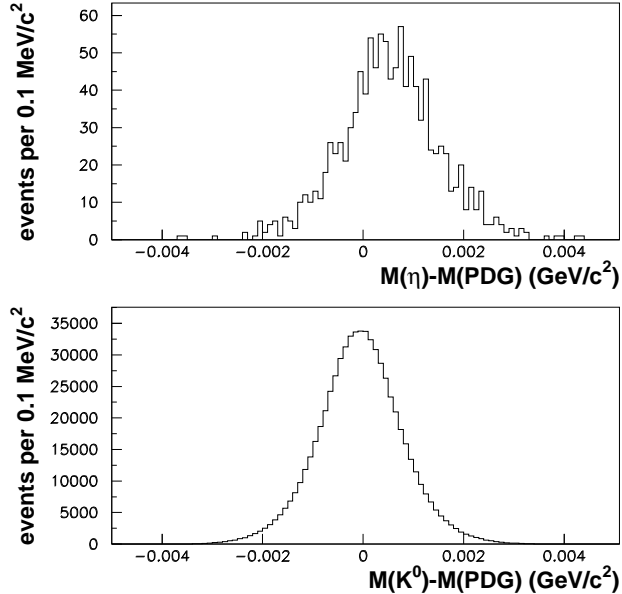}
\caption {Reconstructed masses from 3$\pi^0$ decays, from the $\eta$ sample
(top) and the $K_L$ sample (bottom)}
\label {fig:m3pi0.eps}
\end{center}
\end {figure}

 The distributions of the differences of the 
reconstructed $\eta$ and $K^0$ masses
from the world average masses\footnote{
The values are 0.547300 GeV/c$^2$ for the
$\eta$ mass and 0.497672 GeV/c$^2$ for the $K_L$ mass}
are shown in Figure~\ref{fig:m3pi0.eps}.
The average values of these distributions are 
$550\pm30$ keV/c$^2$ for the $\eta$ sample and $-43\pm1$ keV/c$^2$ 
for the $K_L$,
where the errors quoted are only statistical.
When the same analysis is applied on the Monte-Carlo samples, the
shifts observed between the input mass and the average
reconstructed mass are 
$7\pm5$ keV/c$^2$ for the $\eta$ case and $+4\pm3$ keV/c$^2$ for the
$K_L$ sample. This shows that residual bias from reconstruction
and resolution effects are small.

 The sensitivities to the non-linearities discussed in the previous
section are estimated changing the parameters $\alpha$ and $\beta$ 
in Equation 1 and recomputing the reconstructed masses in the
Monte-Carlo samples. The uncertainties associated with the
allowed range of $\alpha$ and $\beta$ are $\pm$2~keV/c$^2$ each on both
the $K^0$ and $\eta$ masses. This small level of uncertainty
is made possible by the use of symmetric decay configurations.
The uncertainty associated with residual non-uniformity in the
calorimeter response (parameter $\gamma$) is evaluated by comparing
the reconstructed masses with and without the final
correction derived from the 2000 $\pi^0$ data and is found to
be $\pm$23~keV/c$^2$ for both the $\eta$ and $K^0$ masses. This is
essentially equivalent to varying the parameter $\gamma$ within 
$\pm$10$^{-5}$~cm$^{-1}$.

 The reconstruction of the mass is also sensitive to the effect
of energy leakage from one shower to another. The uncertainty
associated with the shower profile parameterisation is taken to be the difference
between the shower profile predicted by GEANT and the profile measured
with electron data. The uncertainty associated with the Monte-Carlo
modelling of the residual reconstruction bias is taken to be half
the size of this bias.
This leads to a systematic error on the mass
of $\pm$33~keV/c$^2$ for the $\eta$ and $\pm$20~keV/c$^2$ for the $K^0$.

 The uncertainty in the photon position measurement is estimated
using the nominal calorimeter geometry without the corrections derived
from $K_{e3}$ data. The corresponding uncertainty in the $\eta$ mass
is $\pm$9~keV/c$^2$, and $\pm$5~keV/c$^2$
on the $K^0$ mass (Because there is no
collimation for the $\eta$ case, the centre of gravity of the events
is spread over the full calorimeter, unlike the $K_L$ case, and this
enhances the sensitivity to this effect).

 Finally, the energy response has non-Gaussian tails arising mostly
from hadron production at the early stages of the shower. To study
the sensitivity to this effect, these tails were parameterised from
$K_{e3}$ and $\pi^0$ data and introduced in the Monte-Carlo. The effect of
using this parameterisation in the Monte-Carlo
is a change of 2~keV/c$^2$ for both the $\eta$ and $K^0$ cases. This
change is taken as the estimate of the uncertainty.

 The effect of accidental activity from the $K_L$ beam has been
investigated and found to be negligible on the reconstructed kaon
mass.

 Combining the various contributions in quadrature, the total systematic
error amounts to $\pm$41~keV/c$^2$ for the $\eta$ mass and $\pm$31~keV/c$^2$
for the $K^0$ mass. Correcting the data for the small observed
Monte-Carlo shifts, the following values are thus obtained:

$$ M_{\eta} = 547.843 \pm 0.030_{stat} \pm 0.005_{MC} \pm 0.041_{syst} \; \mathrm{MeV/c}^2$$
$$ M_{K_L}  = 497.625 \pm 0.001_{stat} \pm 0.003_{MC} \pm 0.031_{syst} \; \mathrm{MeV/c}^2$$

 Many cross-checks of this measurement have been made. The stability
of the reconstructed masses as a function of several variables
has been investigated and found to be within the quoted
systematic errors. 
The same value of the
mass is found for $\eta$ originating from both targets.
Repeating the procedure on the lower statistics samples collected
in 1999 has yielded consistent results.
For the neutral Kaon case a similar
analysis as for 3$\pi^0$ has been performed on 2$\pi^0$ decays, using events
collected in the $\epsilon'/\epsilon$ runs and also, separately, those
obtained at the same time as the $\eta$ decays. The agreement
obtained is particularly significant because we have found that the
effect of the dominant systematic uncertainty, which originates from
energy leakage effects, is actually anticorrelated between the 2$\pi^0$ and
the 3$\pi^0$ samples.

As an alternative to using symmetric decays for the $K^0$ mass
measurement,  events have been selected by an algorithm which requires
minimum sensitivity to energy non-linearities. This gives a higher
statistics sample, spanning  a wider range of energies, and the $K^0$ mass
measurement agrees to better than 10~keV/c$^2$.

 Relaxing the cut on the symmetry of the decay and asking 
$E_{\gamma}$~$>$~6~GeV and at least 25~cm separation between
photons, the reconstructed
masses move by -9~keV/c$^2$ for the $\eta$ and
-10~keV/c$^2$ for the $K^0$, while the total systematic error is increased
by almost a factor two because of the much stronger sensitivity to
energy non-linearities.

 The $K_L$ to $3\pi^0$ decays have also been analysed using a kinematical fit.
In addition to four energy and momentum constraints for the decay,
the three photon pairs were constrained to have
invariant masses equal to the $\pi^0$ mass. The linearised constraints equations
were expressed
in terms of two impact coordinates and energy of each photon in the calorimeter.
In the fitting procedure the $K^0$ mass as well as its energy and
decay vertex position were determined.  
Typically the fit converged after 2-3 iterations.
The measurement errors were slightly adjusted to have good stretch functions
and the resulting
fit probability distribution was flat except for an enhancement below
$5\%$. Using the fitted mass and its error, the $K^0$ mass was determined
from the weighted average in the $\pm 5\,$MeV/c$^2$ mass window around the 
world average
mass. This method has somewhat different sensitivities to systematic
effects than the method discussed previously. The results from the
two methods were compared for various cuts on the event
selection, and they were found to agree within $\pm10$~keV, which is
within the quoted systematic uncertainty.

 Another independent cross-check of systematic effects can be
performed using $\pi^0, \eta \rightarrow \gamma \gamma$ produced
at the targets. Using these two different decays, the ratio
of the $\eta$ mass to the $\pi^0$ mass can be computed. The
selection of the events can be done such as to have
either (a) the same energy for the $\pi^0$ and $\eta$, but different
separation between the photons (by a factor 4 which is the mass
ratio) and thus different systematics in the energy leakage from
one cluster to another, or (b) the same distance between the two
photons but then energies differing by a factor 4 and thus a sensitivity
to the energy non-linearity. Symmetric $\gamma \gamma$ decays are
selected requiring $0.8 < \frac{E_{\gamma}}{E_{tot}} < 1.2$. 
To reduce background, events with energy leakage in the hadron calorimeter
are removed. 
The selected range for the $\eta$ energy is 90-120 GeV. 
In case (a) the same
range is used for the $\pi^0$ events, while in case (b) the energy
range for the $\pi^0$ is 22.5-30 GeV.
The values obtained for the $\eta$ mass are
$M(\eta) = 547.80\pm0.14$ MeV/c$^2$ using the selection (a) and
$M(\eta) = 548.15\pm0.35$ MeV/c$^2$ for selection (b). The errors
quoted include the effect of all the systematic uncertainties discussed
above. In case (a), the dominant uncertainty is the systematic uncertainty
on the correction of energy leakage from one cluster to the other. In
case (b), it is the systematic uncertainty associated with energy
non-linearity.
These measurements suffer from larger sensitivities
to systematic effects than the
one using 3$\pi^0$ decays, but within the uncertainties, there is
good agreement between the results, thus giving a further cross-check
that systematic effects are under control.

\section{Conclusions}
 
 From the 3$\pi^0$ symmetric decay configuration, 
the following
mass values are obtained:
$$ M_{\eta} = 547.843 \pm 0.051 \; \mathrm{MeV/c}^2$$
$$ M_{K^0}  = 497.625 \pm 0.031 \; \mathrm{MeV/c}^2$$

 The $K^0$ mass is consistent with the present world average within 1.1 standard
deviations, with a similar
uncertainty. The uncertainty on the $\eta$ mass is about 2.4
times smaller than the current world average uncertainty, while the eta
mass differs 
from the world average by about 0.1\%, corresponding to
4.2 standard deviations.

\section*{Acknowledgements}
We would like to warmly thank the technical staff of the participating
laboratories and universities for their dedicated effort in the
design, construction, and maintenance of the beam, detector, data
acquisition and processing.

\end{document}